# Societal impacts of big data: challenges and opportunities in Europe

Martí Cuquet [a][*], Guillermo Vega-Gorgojo[b], Hans Lammerant[c], Rachel Finn[d], Umair ul Hassan[e]

[a] *Semantic Technology Institute, University of Innsbruck, Technikerstraße 21a, 6020 Innsbruck, Austria*
[b] *Logic and Intelligent Data, Department of Informatics, University of Oslo, Ole-Johan Dahls Building, Gaustadalléen 23 B, N-0373 Oslo*
[c] *Law, Science, Technology and Society research group, Vrije Universiteit Brussel, Pleinlaan 2, 1050 Brussels, Belgium*
[d] *Trilateral Research, Ltd., Crown House, 72 Hammersmith Road, London, W14 8TH, United Kingdom*
[e] *The Insight Centre for Data Analytics, National University of Ireland, Galway, The DERI Building, IDA Business Park, Lower Dangan, Galway, Ireland.*

**Abstract**

This paper presents the risks and opportunities of big data and the potential social benefits it can bring. The research is based on an analysis of the societal impacts observed in a set of six case studies across different European sectors. These impacts are divided into economic, social and ethical, legal and political impacts, and affect areas such as improved efficiency, innovation and decision making, changing business models, dependency on public funding, participation, equality, discrimination and trust, data protection and intellectual property rights, private and public tensions and losing control to actors abroad. A special focus is given to the risks and opportunities coming from the legal framework and how to counter the negative impacts of big data. Recommendations are presented for four specific legal frameworks: copyright and database protection, protection of trade secrets, privacy and data protection and anti-discrimination. In addition, the potential social benefits of big data are exemplified in six domains: improved decision making and event detection; data-driven innovations and new business models; direct social, environmental and other citizen benefits; citizen participation, transparency and public trust; privacy-aware data practices; and big data for identifying discrimination. Several best practices are suggested to capture these benefits.

**Highlights**

- Conducted six case studies on the societal impacts of big data
- Current legal protection does not scale well and needs to be adapted to big data
- Moving the decision-making to the design phase reduces transaction costs
- Potential benefits in decision making, participation, trust, environment and more
- Need to promote European infrastructures, data partnerships and education policies

---

[*] Corresponding author: marti.cuquet@sti2.at

# 1 Introduction

The digital and connected nature of modern day life has resulted in vast amounts of data being generated by people and organisations alike. This phenomena of an unprecedented growth of information and our ability to collect, process, protect, and exploit it has been described with the catchall term of Big Data. From a technical perspective, Big Data is commonly characterized by its volume, velocity, variety, variability, and veracity (Diebold, 2003; Dumbill, 2013), which each pose a number of technical challenges that need to be addressed (Jagadish et al., 2014). However, the acquisition, analysis, curation, storage and usage of Big Data have also an existing and potential societal impact on both the public and private sector, as well as on citizens. In this direction, several studies have been recently conducted to reflect about the use of big data to analyse social change (Manovich, 2012), analyse the data privacy needs (Cranor et al., 2016), raise concern over social and ethical issues (Boyd and Crawford, 2012) and identify ethical, social and policy challenges (Metcalf et al., 2016).

Big Data has indeed a wide range of implications that can be related to economic, legal, social, ethical and political issues. For example, the use of Big Data for the purpose of business development and organization performance raises economic issues. Privacy and data protection are involved in the legal and ethical aspects of Big Data. In term of social issues, the Big Data plays a role in transparency and social development. The quick and sometimes uncontrolled transfer of data across boundaries raises geopolitical issues. All these issues arise in Big Data practices that have effects in third parties that had no direct involvement in the activity itself. These impacts–positive if the action causes a positive effect or benefit to the third party, negative if it causes cost or harm–arise from decisions, activities or products by stakeholders such as industry, researchers and policy-makers.

The Big data roadmap for cross-disciplinarY community for addressing societal Externalities (BYTE) project was funded by the European Commission to focus on the positive and negative impacts of Big Data in Europe. As part of the project, a set of case studies was conducted in six different sectors (Vega-Gorgojo et al., 2015), including a horizontal analysis (Lammerant et al., 2015a), and roadmaps were developed to provide the necessary research (Cuquet and Fensel, 2016) and policy (Grumbach et al., 2016) steps to tackle such impacts and develop a socially responsible big data economy in Europe. This paper reports on the outcomes on the case studies and the insights gained, analyses the risks and opportunities of big data from a legal perspective, and presents the potential social benefits it can bring.

The rest of this paper is organized as follows. Section 2 describes the research methodology, with a definition and classification of impacts, the design and methods used in the case studies and the methodology of the horizontal analysis. Section 3 summarizes the case studies of Big Data in six sectors and also highlights the impacts within each sector. Section 4 presents a cross-case analysis of negative impacts and how they can be addressed, with a special focus given on risks and opportunities coming from the legal framework. Section 5 underlines the key societal benefits of Big Data as observed through the case studies, exemplified in six different domains. Section 6 concludes the paper with a summary of the results, and suggests several best practices to capture the positive benefits of Big Data.

# 2 Methodology

Within this work, we used the economical concept of externality, which the OECD defines as a "situation when the effect of production or consumption of goods and services imposes costs or benefits on others which are not reflected in the prices charged for the goods and services being provided" (Centre for Co-operation with European Economies in Transition, 1993, p. 44), and broadened it to encompass also social, ethical, legal and political effects. Note that such positive or negative impacts arise from big data practices rather than from big data itself, and include not only actual processes but also future potential activities, opportunities and risks given by the existence of data. A total of 73 potential societal impacts arising from big data practices were identified in order to be considered in the case studies (Donovan et al., 2014; Vega-Gorgojo et al., 2015). A key part in the study of impacts is to identify which are the affected third parties. Here, we broadly classified them as citizens, corporations in the private sector and organisations in the public sector (including governments and other public institutions such as universities). Moreover, as the boundary between inside and outside of an organisation is not always clearly defined, especially when it comes to big data practices, there is a certain arbitrariness regarding which are the affected third parties. Thus, we considered not only effects on third parties but also ones within the same organisation producing them, and adopted the term impact as an enlargement of the externality concept. Impacts were further categorised according to their main topic: business models, data sources and open data, technologies and infrastructures, policies and legal issues, and social and ethical issues.

We aimed to gather evidence about the societal impacts of big data in six different domains: crisis informatics, culture, energy, environment, healthcare and smart cities. Our research agenda focused on the economical, social, ethical, and legal aspects of data. We employed multiple sources of evidence in order to augment the validity of our findings, as recommended in qualitative research (Creswell, 2009, chap. 9) and case study methodologies (Yin, 2014, chap. 4). We thus arranged interviews with data analysts and IT engineers from the organizations involved in the case studies.

In all cases we sent some background information to the interviewees beforehand and gave a brief introduction of our research at the beginning of the interviews. We used a questionnaire that covers our research topics, i.e. economical impacts, legal issues of big data, main risks with respect to policies, social implications of big data, use of personal or private data, data leakage issues, and impact of big data in the relations with other players.

We adopted a semi-structured interview approach letting the interviewees speak and elicit their views and opinions, while we aimed to cover all the topics in our agenda and to request further explanations. After each interview we prepared the transcript, had an internal revision and then shared the report with the interviewee (normally within a week). The length of the interviews ranged from 30 to 90 minutes. Overall, we conducted 39 interviews with data experts from each case study discipline between November 2014 and June 2015.

Besides the interviews, we held 6 disciplinary focus groups with more than 50 big data experts in the corresponding fields, and obtained new insight about the current use of big data and the positive and negative impacts associated with the case studies.

Concerning the data analysis, we followed a procedure based on the qualitative (Creswell, 2009, chap. 9) and case study (Yin, 2014, chap. 5) research literature. In a first step, we prepared the transcripts of the interviews (as explained above) and the focus group sessions. Next, we read through all the transcripts and began the coding of the data –this is the process of organizing the material into chunks of text before bringing meaning to information (Rossman and Rallis, 1998). In particular, we identified specific statements, organized them into categories –corresponding to the topics above– and assigned them labels. Finally, we identified the themes for analysis based on the coding scheme and obtained the findings of the case study.

Next a horizontal analysis was made of the case studies and the societal impacts found. First the big data practices were reviewed. Big data is not a very well defined term and often used as a buzzword. Therefore it was important to take stock of the actual big data practices in the case studies. The use of big data in the case studies was compared along the range of technical challenges (volume, velocity, variety, veracity) and these challenges were mapped along the Big data value chain: data acquisition, data analysis, data curation, data storage and data usage (Curry, 2016). This analysis allowed us to analyse the link between technical challenges and societal impacts. In the second step the societal impacts found were reviewed to check which impacts are present across the sectors and which ones are sector-specific. For the societal impacts that were more generally present an evaluation and analysis was made of best practices to address these impacts, both in terms of capturing the positive benefits and to diminish negative effects. Together with the results of the six case studies, in the horizontal analysis we also use preliminary findings of a seventh study in the transport sector that focuses on the increasing availability and use of data in the maritime industry. These findings show some barriers for adoption of big data that are worth mentioning.

## 3 Societal impacts of big data in the BYTE case studies

In this section we outline the economic, social and ethical, legal, and political impacts found in each of the BYTE case studies, as summarised in Table 1.

### 3.1 Crisis informatics

This case study examines the use of social media data to assist in humanitarian relief efforts during crisis situations (Finn et al., 2015). It evidences positive economic impacts related to the better provision of humanitarian relief services, the provision of better, more targeted and more timely social services and better resource efficiency in providing these services. A significant facet of this is the collection of reliable information, on the ground, much more quickly to aid the situational awareness of the humanitarian organisations. Nevertheless, the integration of big data, or data analytics, within the humanitarian, development and crisis fields has the potential to distract these organisations from their core focus and may represent a drain on scarce resources.

Table 1. Societal impacts grouped as economic, social and ethical, legal, and political areas, and their positive or negative presence in the BYTE case studies.

| | Impacts | Crisis Informatics | Culture | Energy | Environment | Healthcare | Smart Cities |
|---|---|---|---|---|---|---|---|
| ECONOMIC | Improved efficiency | + | | + | + | + | + |
| | Innovation | + | + | + | + | + | + |
| | Changing business models | - | + - | + - | + - | | + - |
| | Employment | | | + | + | | + - |
| | Dependency on public funding | - | - | | - | - | - |
| SOCIAL & ETHICAL | Improved efficiency and innovation | + - | + | + | + - | + - | + |
| | Improved awareness & decision-making | + | | | + | + | + |
| | Participation | + | + | | + | | + |
| | Equality | - | | | | | - |
| | Discrimination | - | | | - | - | |
| | Trust | + - | | - | | + - | + - |
| LEGAL | Privacy | - | - | | - | - | - |
| | IPR | - | - | | - | - | |
| | Liability & Accountability | - | | - | - | - | - |
| POLITICAL | Private, public & non-profit sector | - | - | | - | | - |
| | Losing control to actors abroad | - | - | - | | | - |
| | Improved decision-making & participation | | | | + | + | + |
| | Political abuse & surveillance | - | | | - | | |

As for the social and ethical impacts, improved humanitarian services in and of themselves demonstrate a clear social and ethical gain for society. Further positive impacts in this area include an increased awareness around the need for socially responsible and ethical data practices, the development of tools to ensure ethical data practices, and increased participation through a crowdsourcing model. However, the use of social media data to augment humanitarian relief services raised a number of potential negative impacts such as the misuse of sensitive information, the potential misinterpretation of data and potential for discrimination.

While the use of social media certainly raises significant issues with respect to privacy, data protection and human rights, these issues are central to the way that data is being handled within the organisations in the case study. Our analysis makes clear that experts in this area are committed to ensuring ethical data practices within crisis informatics, although much work remains to be done in relation to the unintentional sharing of sensitive information, the protection of vulnerable individuals and the potential for discrimination that could result from this data processing.

Finally, negative political impacts include a tension between private companies with extensive data analytics capabilities and humanitarian and other relief organisations. Humanitarian organisations are increasingly frustrated with private companies arriving during crises and leaving once the crisis has finished, without sharing or further developing the technological tools and capabilities that they introduced. Further, they are also concerned about being dependent upon them for infrastructure, technological capabilities or other resources, as these organisations have proven to be unreliable partners. In addition, the hosting of data on US soil means that the data becomes subject to US law, which introduces a vulnerability for people contained in those records.

*3.2 Culture*

This case study examines a pan-European cultural heritage organisation that acts as an aggregator of metadata of European cultural heritage organisations. Concerning economic impacts, the main positive effect is the creation of social and cultural value by raising the visibility of the collections and of the providers and driving more traffic to these national and local sites. However, the restrictive funding environment and stakeholders' inability to exploit metadata directly can act as barriers to innovation as well. Indeed, private organisations have not yet found appropriate ways to monetise cultural data or to generate commercially viable applications.

On the societal and ethical side, the overarching impact is associated with the creation and enrichment of cultural heritage data. This is especially valuable to researchers and academics that can use this search facility to gather insights and a more accurate presentation of cultural and historical facts.

Reuse of cultural data is not absolute and for cultural data to be lawfully reused it needs to be done so in accordance with relevant legal frameworks. In fact, managing intellectual property issues that arise in relation to the re-use of cultural data is perhaps the biggest challenge facing big cultural data driven initiatives. Further, arranging the necessary licensing agreements to enable reuse of cultural data can be arduous, especially as there is limited understanding and information about how rights statements and licensing frameworks can support stakeholders in capturing the full value of the data. In addition, data protection was identified as a legal barrier to some models that incorporate the use of cultural interaction data, and for also limiting the reuse of certain forms of cultural data, such as data including references to sensitive personal material.

Finally, political issues arise in relation to making the data open because it can lead to a perceived loss of control of data held by national institutions thereby causing intra-national tensions. This tension is also fuelled by reluctance on part of institutions to provide unrestricted access to their metadata under a CC0 license (this is a fundamental requirement of this case study and all data partners are required to sign a data agreement to that effect). There also exists a political tension around the American dominance over infrastructure. This has prompted a general trend towards reversing outsourcing practices, and developing infrastructure and tools in-house.

*3.3 Energy*

The energy case study is focused on the impact of big data in exploration and production of oil and gas in the Norwegian Continental Shelf (Vega-Gorgojo et al., 2016). While big data still needs to prove its effectiveness in oil and gas, the industry is beginning to realize its potential and there are many on-going initiatives, especially in operations. In this case study a number of economical impacts associated with the use of big data were identified: data generation and data analytics business models are beginning to get traction, there is a number of commercial partnerships around data and the Norwegian regulator has embraced open data in order to spur competition among oil operators. However, companies are still reluctant to share their data, despite some emerging initiatives. Moreover, existing business models have to be reworked in order to promote the adoption of big data. In addition, the oil and gas industry is becoming interested in hiring data analysts to exploit the potential of big data for the integration of large data volumes, to reduce operating costs and improve recovery rates and to better support decision management.

As for the social and ethical impacts, personal privacy is not a big concern and there is little value of social media. In contrast, some datasets are highly secret and confidential, so cyber-security measures are quite important and have been adopted through the whole industry. Due to the environment concerns of petroleum activities, safety requirements are pivotal and big data can help to reduce environmental impacts by the early detections of incidents, e.g. oil leakages, and by improving equipment efficiency, e.g. through condition-based maintenance. There is also a trust issue with data coming from uncontrolled sources. This is especially relevant when aggregating data or when applying data-driven models.

Petroleum activities in the Norwegian Continental Shelf rely on a mature regulation framework that enforces the separation of policy, regulatory and commercial functions. While production and seismic data are highly regulated by the authorities, other datasets, e.g. operations data, are normally regulated by the terms and conditions of a contract. Indeed, industry stakeholders are becoming increasingly aware of the value of data, so ownership of data will possibly be subject of contention.

Concerning political impacts, operators and suppliers in the oil business are normally international organisations with businesses in many countries. Oil operators purchase data (especially seismic) from other companies with a strong presence in the surrounding areas in order to carry out exploration and scouting activities. Data is thus becoming a valuable asset that is traded internationally. Indeed, some vendors are becoming leaders in big data, and the rest should embrace big data in order to succeed in the future.

*3.4 Environment*

The environment case study has been conducted in the context of an earth observation data portal (EarthObvs), a global-scale initiative for better understanding and controlling the environment, to benefit society through better-informed decision-making. On the positive side, the use of big data is credited with having strong implications on the economic growth, e.g. sea data for fishing nations and weather data for tourism, for the mere direct effects on the IT sector such as opportunities for infrastructure/hardware/data centres, and rent-sharing possibilities for computing. However, big data could as well be seen as a threat by traditional services, for example in the weather forecasting sector.

Our analysis of social and ethical impacts shows the effectiveness of data-intensive approaches in improving the governance of environmental challenges, supporting safe and environment-friendly operations. This has implications on the robustness of the environment to recover after stressing events (resilience), especially in urban context, as well as on the actual quality of life and human health. On the other hand, big-brother-effect and manipulation, real or perceived, can be problematic. In fact, as human society is an integral part of the environment, especially in urban context, the fear of data abuse, privacy violation and the like, may hamper participation and engagement. More subtly, also excessive trust in data-intensive applications has been highlighted as a possible negative implication, in that it would encourage the false believe that the dynamics of the environment can be captured quantitatively, overlooking qualitative aspects that, instead, remain fundamental to comprehend it.

With respect to legal impacts, the growing reliance on data in the environment sector is certainly highlighting many shortcomings of the current legal frameworks, e.g. on IPR, privacy, and authorisation to use the data. Potential problems are more prominent when considering different legislations such as the principle of full and open exchange of data that is simply inconsistent with some of the current national policies. Legal support for citizens in data-related issues is arguably going to become a public service, in the future. Besides, incompatibilities in the legal frameworks in different countries are seen as inhibitors that need to be adapted, in order to remove legal barriers.

Finally, political impacts include increased transparency and accountability, since choices will have to be based on measurable and observable indicators. However, there is a risk of depending on external sources, particularly provided by big players. In this regard, the environment community is concerned about the imposition of commercial products as de facto standards, causing an implicit vendor lock-in for their maintenance and evolution. The use of EarthObvsgraphical data is a potential source of political tensions, e.g. with regards to disputed or otherwise sensitive areas. As Europe leads the innovation in the EarthObvsspatial sector, the shift to a more intensive use of data in the environment sector may put Europe in a primary role on the EarthObvspolitical scene.

*3.5 Healthcare*

This case study focuses on the use of genetic data as it is utilised by a public health data driven research organisation. Overall, this case study highlights a number of positive societal impacts that flow from genetic research and rare gene identification, which is facilitated through the utilisation of big health data. Concerning economical impacts, one important result is cost saving for healthcare organisations that are gained through more accurate and timely diagnoses and efficient treatments. This is particularly important when dealing with rare genetic disorders that may not otherwise attract the attention that disorders and health issues affecting the wider population do. In addition, the utilisation of big data in healthcare produces another economic impact in that it potentially generates revenue especially through the development of marketable treatments and therapies, and the innovation of health data technologies and tools. Nevertheless, research initiatives, such as the institute of the case study, are naturally subject to financial restrictions and cost savings measures implemented by governments.

On the societal and ethical side, the identification of rare genetic disorders provides treatment opportunities for the patient, as well as more effective diagnostic testing for future patients and a greater understanding of rare genetic disorders generally. In addition, analyses of genetic data enable treating clinicians to provide a range of other healthcare services for family members, including genetic counselling. Beyond the initial purpose for the data collection, there is limited or no reuse of genetic data due to legal and ethical issues. However, data reuse may become a stronger focus in the future for the purpose of producing additional benefits for patients and society. There will likely be an increase in focus on research for the purpose of developing personalised medicine treatments, which focuses on improved treatment based on patient drug metabolism. Whilst the positive social impacts of reusing genetic data are obvious, ethical considerations will remain at the forefront of any policies supporting the repurposing of genetic data, especially as it is sensitive personal data. There are potential negative impacts linked to the utilisation of big data in healthcare such as over-medicalization of an otherwise healthy population; discrimination based on the stratification on genotype or in relation to health insurance policies; and incidental findings discovered when data is analysed. For example, cancer genes may be discovered alongside the identification of rare genes or other genetic mutations.

Since health data is by its very nature sensitive data, this case study highlights legal risks associated with data protection and data security. In particular, anonymisation is a legal requirement that is challenging to achieve with genetic data. The issue of data protection and information privacy calls for the development of adequate protections that balance the right to personal data protection whilst fostering innovation in the digital economy. Although data security preservation measures ensure compliance with standard data protection requirements, they can also hinder further research, which in turn, could lead to new developments and treatments. Lastly, threats to intellectual property rights can arise in relation to subsequent uses of big health data, such as in relation gene patenting (and licensing) of new drug therapies, or if it were to be included in works protected by copyright.

Concerning political impacts, the availability of big amounts of data will enable politicians to have more information about different situations in the health sector and thus a better understanding that may lead to improve their decision-making and increases the investments in healthcare.

*3.6 Smart cities*

This case study is focused on the creation of value from potentially massive amounts of urban data that emerges through the digitalised interaction of a city's users, i.e. of citizens and businesses, with the urban infrastructure. The economies of digital, especially big data, favour monopolistic structures, which may pose a threat to the many SMEs in cities and the small and medium cities. However, open source and open platforms, open data, and even open algorithms crystallise as a shortcut and "technology-driven form of liberalisation" accompanying the big data movement in cities, which has the potential to level the playing field and even spur more creativity and innovation. Nonetheless, monetisation of big data remains a moving target. Investments by the private sector need to be secured by reduction or compensation of future uncertainties, otherwise the current state of lack of investments will remain. Similarly, with the increasing potential of machines that learn, old jobs consisting of simple tasks are also at risk. New skills are required and will create new jobs, but numbers are most likely not equal.

As for the social and ethical impacts, the potential of big data to be used for social good is immense, especially in the digitalising city. However, the reliance on data-driven services needs a debate on how we can assure enough equality when there are so many different reasons why not all citizens will reap value from data in equal amounts. This may be due to the digital divide we have been aware of for a while now, or to entirely new challenges through recent technological breakthroughs such as deep learning, which enable machines to learn and take over simple tasks, such as counting cars or recognising letters. Moreover, trust in computing methods for big data may be harder to establish if their rationale cannot be easily explained. Furthermore, businesses, critical infrastructures, and lives may rely on big data algorithms.

Since new sources of data create new ways that data can be misused (including sensitive personal data), this case study questions the suitability of the current legal framework based on data ownership. Indeed, the shift towards big data may imply that data as well as algorithms to mine the data are required commodities to create value through user experience and services.

Finally, political impacts include losing control to big data actors outside Europe. Indeed, big data businesses can improve the European economy, but require a unified European data economy with accordingly unified policies.

**4 Addressing the negative impacts of Big Data**

Big data is often presented as disrupting existing business practices and social relations. Legal frameworks regulating these relations also get affected. They lose their efficacy or get criticised as out-dated and presenting a barrier for the adoption of new technologies. In other words, legal problems show up when there is a mismatch between the regulating frameworks and the new big data practices.

To better understand how legal problems arise from big data practices, we need to look at how these practices affect the interactions between different actors, be they individual persons or organisations. When the legal frameworks regulating these interactions cannot adapt to the qualitative changes in these interactions brought by big data, this will show up as a legal impact (Lammerant, 2016).

Big data is made possible by new developments in distributed computing like cloud technology, which enables processing of very large amounts of data at much higher speed. However, big data cannot be equated with these technologies or cannot be limited to these aspects of volume and velocity. It also implies qualitative changes in terms of what can be done with data: a variety of structured and unstructured data sources can be much easily linked and analysed in new ways. New business models are built upon the capacity to capture value from data through the development of a data value chain along which data is transformed into actionable knowledge. The incentive for big data is that it provides larger efficiency through more fine-grained and targeted control. It can also generate new visibilities of certain phenomena and makes possible new innovative business processes. New data value chains imply new data flows, and new or altered interactions, between actors. Analysing the qualitative changes in these interactions allows us to understand how problems arise and to evaluate the effectiveness of legal frameworks.

*4.1 How big data affects interactions between actors*

The capacity to collect, process and analyse data on a much larger scale has turned data into a network good (Blanke, 2014; Rouvroy, 2016). The growing availability of data results in positive network effects. This means that, when new data becomes available, also new value can be created by re-using older data sources in combination with these new data. A growing amount of data enables to make more value out of data by multiplying its uses or by making possible more fine-grained analysis. These positive network effects are an incentive to enlarge data collection and for business models that are more intrusive towards potential data sources and which demand more interaction with other actors.

These larger data flows result in a higher visibility, and therefore a more in-depth knowledge, of an actor. Big data enables to get a more fine-grained perception of an actor's behaviour and makes possible faster and more targeted reactions towards it. This has an impact on organisational boundaries, which get penetrated by these data flows and become more difficult to uphold. Boundaries are a tool with which an actor maintains its identity and autonomy. From this identity follows an actor's interests, and its capacity for autonomous action enables it to pursue and protect these interests. Legal means like property rights, organisational structures mapping which activities are done by whom and who has access to which information, or technical means like the walls of a house define boundaries between the outside environment and the inside of an organisation or the private sphere of an individual. An actor uses such means to filter what is seen and known about him by the outside world and to keep others out. Impacts on these boundaries also impact the processes through which an actor preserves its identity and autonomy. On the boundary of an organisation gatekeepers or gatekeeping functions are situated, which open and close channels of communication and information flows (Bekkers, 2005). The growing amount and changing characteristics of interactions means that such gatekeepers can become dysfunctional. Big data makes an organisational system more open and therefore gatekeeping becomes more complex. Big data enlarges interdependencies, which can be reciprocal but also very unbalanced, and opens an actor to outside power and control. Efforts to limit interactions clearly conflict with the augmented interaction linked with big data-related business processes. Legal regulation reflects current gatekeeping practices and is therefore also vulnerable to the impact of big data. It has to adapt to remain effective when the organisational boundaries are continuously penetrated with data flows.

Data-driven business processes lead to a shift in the structure of transactions from an exchange of goods towards a delivery of services. An exchange of goods involves a limited amount of information exchange between actors, while the whole transaction takes place during a specific moment or limited time interval. However, the delivery of services is generally more continuous or regular in time and stretches over longer periods. It often also involves a flow of information during the whole period of service delivery. E.g. the practice of condition-based maintenance turns acquiring equipment into acquiring also a service, where the equipment provider signals when maintenance is needed based on a continuous data stream during operations. Developers of software, like operating systems or browsers, use this data to monitor the functioning of their product and provide maintenance through updates. This change in transactions has an important impact on the gatekeeping practices surrounding these transactions. Where the momentary or time-bounded character of a sale naturally limited the data flow involved and the need for gatekeeping, such gatekeeping practices have to adapt to remain effective when the data flow becomes continuous. Maintaining organisational boundaries has to change from a time-bound control of a transaction towards the monitoring and regulation of continuous interactions.

*4.2 Consequences of changing interactions and the appearance of negative impacts*

The negative impacts of big data can be explained from the difficulties to adapt regulatory mechanisms to the new characteristics of interactions and to ensure their effectiveness. The enlarged visibility and penetration of boundaries appear as privacy problems for individuals. As mentioned above, the positive network effects of data are an incentive for business models which are more data-intensive, but thereby also more intrusive towards individuals. E.g. targeted advertising becomes more efficient the more data it can collect, turning personal data into a commercial asset and providing an incentive for more intrusions of privacy. The positive network effects of data also make protection techniques like anonymisation more vulnerable and unreliable on the long term. These network effects can also lead to a propagation of discriminative effects present in data and result in unintended discrimination in other contexts. Decision-making can become more opaque when based on combining a wide range of data sources and thereby less accountable.

Power relations between commercial actors and individuals can get upset. For both individuals and organisations this can lead to mistrust, uncertainty and a fear to lose control and autonomy. In such cases withdrawal and reluctance to participate in big data practices are a rational strategy to reduce uncertainty. Restoring certainty through other means, like adapted legal frameworks, would enable these actors to participate again. The shift to services and the continuous interaction involved generate the need for new forms of gatekeeping and upholding boundaries.

Current legal practices still reflect the perspective of control of a limited amount of interactions. Often they are not able to scale towards more interactive and continuous data flows, due to the high transaction costs involved. E.g. the consent mechanism in data protection law prevents an aggregated treatment and presents a high transaction cost. Therefore it becomes an important barrier whenever a large amount of interactions takes place. More efficient and scalable mechanisms have to be adopted which reduce the transaction costs involved.

*4.3 Addressing the negative impacts of big data*

The negative impacts of big data can be explained from the difficulties to adapt regulatory mechanisms to the new characteristics of interactions. Based on this analysis, an approach can be developed to evaluate and to address the impact of big data on legal frameworks.

First, the objectives of the legal framework need to be clarified and used as a yardstick to check if the balance struck between different objectives and interests in the current practice or legal framework is still delivering an optimal result. If not, it will need adapting. Such evaluation can lead to different conclusions depending on these objectives. E.g. the copyright framework has an economic objective, balancing different interests in order to obtain an optimal result in terms of stimulating innovation on societal level. The copyright framework can be adapted to ensure remuneration for who created the data, but it is possible that limiting this protection provides a better result as it allows for a market expansion due to the network effects of the availability of data. Other legal frameworks deal with several objectives, which cannot be translated in one overall measure. This makes evaluation more difficult as it allows a range of best solutions depending on how the objectives are valued. E.g. the data protection framework has to balance the objective of privacy with economic objectives. An important question in this context is evaluating if the framework can still be applied and retains its efficacy to achieve its objectives.

Second, legal frameworks also provide actors with instruments for gatekeeping. The analysis above made clear that legal mechanisms based on individual transactions or individual control lead to high transaction costs in a context of a larger amount of interactions. Or they become dysfunctional, or they present barriers to big data practices. They need to be substituted for collective or aggregate mechanisms of control and decision-making. This implies a shift from a transaction or an individual control model to a liability model. In the transaction or individual control model the individual decides about each interaction. On the contrary, in a liability model such decision-making during operations happens more collectively or aggregated and the active role of the individual gets reduced to receiving compensation or obtaining a claim towards the big data operator.

Third, a specific method to reduce transaction costs is to move a large amount of the decision-making to the design phase and to create standardised solutions. This implies that legal or social objectives get translated into technical objectives and requirements and get integrated in the design methodologies. Standardisation is then a further collective process creating well-understood common solutions. This reduces the needed information processing in individual decision-making and thereby also transaction costs.

Fourth, we are concerned with socio-technical systems and problems are not purely technical or social. Best practices are often a mixture of legal, organisational and technical elements. Where the development of design objectives and methodologies and standardisation efforts is now mostly focussed on technical elements, this should be broadened to include also legal and organisational elements.

| | |
|---|---|
| ***Copyright & Database Protection*** | Drop the sui generis-right on databases. |
| | Add exceptions for data mining in copyright. |
| | Collective licensing. |
| | Limiting the extent of copyright. |
| ***Trade Secrets*** | Adoption of a legal protection of trade secrets. |
| | Develop standardised solutions. |
| | Toolbox to fine-tune data flows. |
| ***Privacy & Data Protection*** | Promote development of standard evaluation. |
| | Broaden privacy-by-desig for legal and organisation safeguards. |
| | Include aggregated and collective mechanisms in data protection frameworks. |
| | Strengthened role for data protection authorities. |
| ***Anti-discrimination*** | Promote an anti-discrimination-by-design approach. |
| | Develop a transparency and accountability framework. |
| | Coordination and cooperation between data protection authorities and equality bodies. |
| | Engage in a prospective societal and political debate on 'new discriminations'. |

Figure 1. Summary of recommendations for addressing negative impacts of Big Data.

*4.4 Recommendations for specific legal frameworks*

In our research we evaluated several legal frameworks according to this general approach and came to the following conclusions. The following recommendations are summarised in Figure 1.

4.4.1 Copyright and database protection

Evaluations of the current copyright and database protection laws show that this framework is too restrictive. Possible solutions involving legal change are to drop the sui generis-right on databases and to add new exceptions for data and text mining in copyright. A solution within the current legal framework is collective licensing, which lowers transaction costs but preserves remuneration. Extended collective licenses can make this framework more flexible. Still, the evaluations show that limiting the extent of copyright is preferable, as it leads to market expansion due to network effects.

4.4.2 Trade secrets

The case studies showed that companies can have 'privacy' problems as well, in the form of a need to protect confidential information. To address these concerns, the adoption of a legal protection of trade secrets is recommended. Within such a framework stakeholders and standard bodies have to develop standardised solutions and a toolbox of legal, organisational and technical means which can be used to fine-tune data flows.

4.4.3 Privacy and data protection

The principles of the data protection framework can be implemented in a risk-based approach without limiting the extent of their protection. But in such a risk-based application, it is important to evaluate the actual risks associated with the overall set-up of technical, legal and organisational safeguards, instead of assessing the theoretical risks associated with anonymisation techniques or other technical safeguards. To make such overall evaluation possible, the development of standard solutions with known capabilities and risks for the whole range of technical, legal and organisational safeguards needs to be promoted. Further, the privacy-by-design approach needs to be further developed and broadened from a purely technical approach to include also legal and organisation safeguards. Lastly, individual transaction-based elements of the data protection framework, like consent, need to be substituted by more aggregated and collective mechanisms, which includes a strengthened role for data protection authorities.

4.4.4 Anti-discrimination

An important negative impact is the risk for discrimination. Data mining can result in unintended discrimination due to choices in the problem definition and to discriminatory bias in the training data. Effective methods to deal with such bias have recently been developed and are object of further research. These methods can be integrated in auditing tools and in an anti-discrimination-by-design approach, similar as in privacy-by-design. The legal framework concerning anti-discrimination is less developed than data protection and mostly focuses on legal redress, but also establishes equality bodies working on mainstreaming of policies. They can play a role in addressing anti-discrimination in big data as part of these mainstreaming efforts. First by promoting the development of an anti-discrimination-by-design approach. Second by developing a transparency and accountability framework based on both the anti-discrimination legislation and the data protection framework. Coordination and cooperation between data protection authorities and equality bodies on this matter would be very useful. Lastly, big data creates new visibilities and makes it possible to discern between people on a whole range of health- and behaviour-related and other personal aspects. This also provides grounds for 'new discriminations'. The equality bodies will have to engage in a prospective societal and political debate on how to deal with these 'new discriminations'. This will have to lead to the development of new norms on which decision-making is acceptable based on these new visibilities.

# 5 Social benefits of big data

The use of large data sets for data analytics, predictive analytics and deep learning does not only pose legal challenges and opportunities, as discussed in the previous section, but also carries significant potential societal opportunities when data is used responsibly. Many initiatives specifically aim to use the analysis of large data sets for "social good". For example, the Global Pulse initiative seeks to maximise situational awareness in development and humanitarian work in order to produce socio-economic benefits (UN Global Pulse, 2012). The Australian Public Service Big Data Strategy (Commonwealth of Australia, 2013, p. 5) intends to support the use of data in the public sector to reform and improve services, as well as protect citizen privacy. In many cases, in policy circles, business sectors and beyond, data is increasingly being seen as an asset or a resource that can improve business practices, social welfare and citizen trust. Nevertheless, these benefits are predicated on addressing many of the negative impact associated with big data. Despite this challenge, BYTE has identified six areas where the use of big data can result in societal benefit, with their impact on BYTE sectors depicted in Figure 2. These six areas are improved decision making and event detection, including efficient resource allocation; data-driven innovations, including new business models; direct social, environmental and other citizen benefits; citizen participation, transparency and public trust; privacy-aware data practices; and big data for identifying discrimination.

## 5.1 Improved decision making and event detection, including efficient resource allocation

Stakeholders from across the BYTE case studies noted that one of the key societal benefits of the use of data analytics for large volumes of data was improvement in decision-making and situational awareness. This also includes improvements in efficiency of resource allocation based on a better understanding of the situation and an ability to target resources. This improved decision-making and event detection was evident in almost all of the BYTE case studies.

In relation to crisis informatics, this included a better overview understanding of the ways in which particular citizen groups or regions were affected by crises. Social media analytics and other data analytics tools were used to establish the "who, what and where" of a crisis more quickly and to predict where specific resources might be needed according to an International Government Organisation representative. Many stakeholders noted that better data collection and analysis would enable humanitarian organisations to "provide relief faster", "allocate resources where need is greatest" and target services, especially using information that is provided directly by the public. Furthermore, the movement from Web 2.0 sharing and predictive analytics was also thought to carry the potential to improve crisis preparedness, for example, by allowing a pre-crisis distribution of resources to areas where populations might move in the event of a crisis.

In the oil and gas case study, improved decision-making took the form of identifying and responding to equipment failure. One of the case study organisations, Soil, reported that using sensors to compare pre-drilling information about the seabed with current readings would allow them to identify failures and respond more quickly if "something is going wrong". In emergencies, this data can be shared with other relevant organisations to address the problem as quickly as possible.

In the environment case study, improved decision making and response was focused on environmental and public health issues. This included progressing towards more environmental sustainability, improved safety and reduced disaster risks. Specifically, environmental data could be used to feed directly into policy-making, enabling better informed decision-making by legislators and greater environmental impacts of policy.

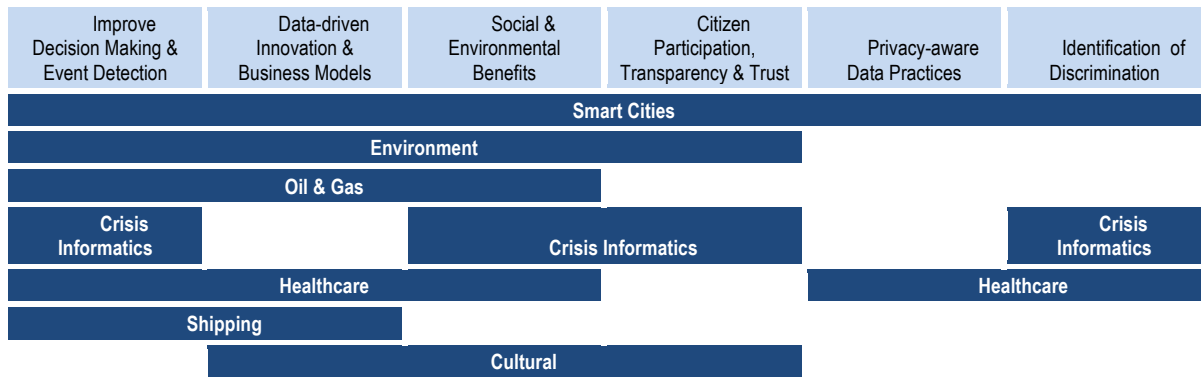

Figure 2. Mapping of the social benefits of Big Data against different sectors.

In healthcare, genetic data was used to inform decision making related to treatment. This included early diagnosis and thus improved and better targeted treatment regimes. One specific benefit in the genetic sphere was the possibility to treat conditions before they emerged, therefore essentially preventing diseases from occurring. In other contexts, it meant sharing information to identify best practices for particular diseases to enable better decision-making by other medical professionals.

In the shipping and transport case study, improvements in decision-making included better optimisation of business and technical processes. For example, this could take the form of design improvements in equipment or processes. According to case study informants, this meant better allocation of scarce resources that theoretically could be used for other business purposes.

Finally, in relation to smart cities, this benefit was evident in relation to incident response or policy-making. In one example, police and fire services could respond to incidents more effectively using information from various sources such as surveillance cameras, GPS equipment on response vehicles, fine and smoke sensors as well as others. This would give them a better situational picture that would also enable an efficient deployment of forces. In relation to policy-making, this included the use of data to improve urban planning, including allocating resources in particularly needy areas and also analysing behavioural changes resulting from those interventions.

*5.2 Data driven innovations, including new business models*

Another direct societal benefit of the proliferation and popularisation of big data analytics is the use of data to drive innovation, including identifying new insights, driving cultural innovation or supporting new business models. While there is overlap here with some economic impacts, the effects of these practices are often realised socially as well as economically. These impacts were evident in the cultural, environmental, oil and gas, shipping/transport and smart cities case studies.

In the cultural case study, open data was used to drive innovation in both socio-economic and cultural terms. With respect to socio-economic innovation, the cultural data case study focused on using open data to create new products and services, including generating efficiencies or new insights by utilising open data from catalogues like Datahub (http://datahub.io) – as well as commercial data providers like Socrata (http://socrata.com) or Factual (http://factual.com). Cultural innovation primarily referred to creating and enriching cultural resources to generate social value for citizens. This could be achieved by facilitating access to cultural data and thus providing richer data sets that make data sharing more attractive. This could result in a positive feedback loop that encourages others to make their cultural data more available, thus facilitating even further cultural innovation.

The environmental data case study discussed the use of diverse data sets to generate new insights. This came in the form of using high-resolution satellite data alongside building information models to assess building vulnerabilities and prevent damage before it can occur. Innovative services such as these are a direct result of combining open and proprietary data.

Both the oil and gas and the shipping case studies focused on new business models as a form of data driven innovations. Like the environmental case study, this was focused on new service offerings. In oil and gas, this included data collection and supply as a service, as well as data storage and analysis as services. The latter was especially relevant when historical data had already been collected, and data from new sensors could be used to compare current models with historical ones.

Finally, in relation to smart cities, data driven innovation was associated with the cyclical investment in data platforms and sharing. Specifically, open source platforms and infrastructure funded by specific investment programmes were being used to share data, which led to data infrastructures becoming quasi-utilities that served citizens. As such, cities were able to improve services and provide an ignition platform for new businesses and new service offerings.

*5.3 Direct social, environmental and other citizen benefits*

Innovative practices surrounding big data may also result in social and ethical benefits for citizens and the environment, either through improved services, better situational awareness or personalised or targeted services. These benefits ranged from environmental protection to improved life chances to personalised government services for individual citizens. These benefits were seen across the crisis, environmental, oil and gas, health and smart cities case studies.

Within the crisis management case study, direct benefits were reported in relation to humanitarian aid, safety and survival. In crisis situations, improved information, situational awareness and resource allocation all lead to operations that have direct impacts on citizen's survival and health during these situations. Furthermore, indirectly, better resource allocation means that humanitarian practitioners can reach and provide aid to greater numbers of people, thus having an impact on larger populations.

The environmental data case study found that these benefits were primarily visible in relation to environmental and public health issues. Environmental impacts included better planning operations to ensure that they are safer and more environmentally friendly. In relation to public health, it also included better health and education services resulting from better data sharing and analysis as well as better handling of the environmental causes of ill-health, for example through pollution reduction measures.

In oil and gas, these benefits primarily focused on safety and reduced environmental impact. Big data can be used to improve safety when it is being used to monitor reservoirs. In addition, it can also give a clearer picture of operations and enable the detection of oil leakages or equipment damage that could lead to an incident. Thus, using data analytics for predictive maintenance or early identification of incidents will reduce the environmental impact of industrial operations.

The healthcare case study found direct citizen benefits in relation to personalised treatment methodologies. The genetic elements of the data analytics being undertaken within the health case study made it possible to design and implement personalised medical interventions. These had knock-on impacts on resource allocation and public and private sector integration for service delivery. In these cases, both patients and service providers benefited from more relevant information for the patient and better outcomes in relation to service delivery.

Citizen benefits in smart cities included benefits related to citizen experience and welfare. For example, better access to more targeted and relevant public services would improve both service delivery for public authorities but also result in a better experience for citizens. Better citizen's experience also included quality of life issues like "getting from A to B on time". Furthermore, like crisis and health, better resource efficiency through targeted services would also have knock-on benefits for citizens more generally as more resources were available for public service delivery, allowing public authorities to reach greater numbers of citizens.

*5.4 Citizen participation, transparency and public trust*

Social benefits of big data analytics also emerged in relation to enabling citizen participation in data collection and analysis as well as greater transparency and public trust. This was principally evident in the environmental data and smart cities case studies.

With respect to environmental data, citizen science initiatives as well as open data portals were instrumental in involving citizens and increasing transparency and trust. There were significant opportunities for crowd sourcing of environmental data collection within this case study, which resulted in a democratisation of knowledge and greater transparency in the final results. In addition, open data portals related to climate data, for example, provide the public with an opportunity to check whether policy-makers are adequately responding to environmental threats, thus making politics more accountable.

Crowdsourcing was also used in the smart cities case study. In this case, sensor data was combined with citizen-provided data through smart phones and specialised software to design services. It was also used to "tap into" citizen led innovations and co-creation or collaborative policymaking or service delivery, including through public-private partnerships.

*5.5 Privacy aware data practices*

Despite the fact that privacy is often constructed as a challenge to innovation (Michael and Miller, 2013), the BYTE research found that in some case studies one of the unexpected societal benefits of big data practice has been increased attention to privacy and data protection issues by practitioners. For example, practitioners in crisis informatics, environmental science and healthcare have been particularly sensitive to potential privacy issues and have constructed good practice strategies to both protect privacy and provide innovative, data-dependent services.

In crisis informatics, they have developed tools, standards and procedures to ensure ethical data practices. Specifically, they rely on the International Committee of the Red Cross's (ICRC) data protection protocols to guide the collection of data about people, including especially information from social media. They also reduce the granularity of the data so that Twitter handles, tweets and any other personal information is not visible in the public outputs, just aggregated data. Finally, these practitioners are also using risk assessments that include privacy considerations to ensure that the data of people providing information is adequately protected.

In the environment case study, there is a reported increase in awareness about privacy and ethical issues as they relate to big data. In response, practitioners have developed better self-assessment profiles and tools, and have provided communities with templates to enhance their awareness about privacy issues. Each of these has contributed to more responsible data practices.

Finally, practitioners in the healthcare case study have been working with patient data and confidentiality for some time, and have transferred lessons already learned to data analytics. This has included anonymising data at the software level, though adding computer-programming expertise to their ecosystem. They also rely on anonymisation practices already in place within the diagnostic labs to ensure the confidentiality of patient data.

*5.6 Big data for identifying discrimination*

Finally, despite the serious and significant potential for big data to result in discrimination, the BYTE project has also found that in certain circumstances, big data can be used to identify and, consequently, combat discriminatory practices. Systems like DCUBE (Pedreschi et al., 2013; Ruggieri et al., 2010) can be used to identify discriminatory classification rules from the historical data in order to intervene in these practices. Other pre- and post-processing techniques can also be used to remove or compensate for discrimination within training datasets, including massaging the data, reweighting particular variables, resampling or applying model correction methods. Each of these methods can both combat discrimination as well as ensure responsible and ethical data practices moving forward within the big data landscape.

**6 Concluding remarks**

In this document, we have revisited the impacts found in the BYTE case studies, the effect of big data in legal frameworks and the benefits that big data can deliver to society. In all case studies we have observed positive impacts due to improved efficiency and innovation, both from an economic and a social and ethical point of view. Economically, this is related as well to the change and new appearance of business models. This is perceived as a negative impact when it leads to the dominance and dependence of a few big players, but on the other hand it also provides opportunities for niche players. New data-driven innovations and business models have also been observed as a positive social impact, especially in the cultural, environmental, oil and gas, shipping/transport and smart cities case studies. In addition, public funding is viewed as an important requirement to start the data economy, and in particular to address the creation of data platforms and data sharing. The positive social impact of improved, evidence-based decision making is present in several case studies. However, there are also concerns on the representativity and possible bias of the data or models being used. This and other trust concerns show up as negative impacts in most of the sectors, thus potentially diminishing the effect of positive impact. The same applies to equality and discrimination impact.

From a legal point of view, data protection and intellectual property rights are seen as important tools to protect economic and social values from negative impacts. We have presented how big data practices affect the interactions between actors and the need to adapt the legal frameworks to avoid negative impacts. It has been suggested that the objectives of legal frameworks have to be clarified and evaluated to see if the framework is still effective. Moreover, legal mechanisms based on individual transactions or control need to be substituted with collective or aggregate mechanisms, and as much as possible of the decision-making has to be moved to the design phase, translating legal and social objectives into technical ones, and widening the standardisation efforts to legal and organisational elements, and not only to technical ones. We have also presented recommendations for four specific legal frameworks, namely copyright and database protection, protection of trade secrets, privacy and data protection and anti-discrimination.

We have also identified six areas with a potential social benefit from the use of big data common to all or almost all sectors considered. Already mentioned are data-driven innovations and business models, and the use of data analytics for large volumes of data to improve event detection, situational awareness, and decision making to e.g. allocate resources efficiently. Another area is that of better environmental protection and efficiency and direct social impact to citizens through e.g. individual targeted services. Big data can also enable citizen participation and increase transparency and public trust, although this will require efforts to develop data skills among the general public. An unexpected social benefit related to privacy was found in the increased attention paid to privacy and data protection by big data practitioners. Finally, despite the negative impact big data can have on discrimination, the BYTE project has also found that big data also gives the means to identify and combat such practices.

In order to capture these benefits, several best practices have been suggested by the BYTE project (Lammerant et al., 2015b) and further addressed in the research (Cuquet and Fensel, 2016) and policy roadmaps (Grumbach et al., 2016). They involve public investments and funding programs to solve the scarcity of European big data infrastructures, promote research and innovation in big data, open more government data and persuade big private actors to release some of their data as well, so data partnerships can be built around them. New data sources and business models also need to be promoted. Interoperability has also been shown to be a key enabling factor. In addition, education policies have to address both the current scarcity of data scientists and engineers, but also the inclusion of data skills in general educational programs.

To address discrimination, equality and trust, privacy-by-design methods should be extended to anti-discrimination-by-design and analogous approaches, and transparency and new accountability frameworks need to be based both on legislation and on a data protection framework. Overall, policy makers, regulators and stakeholders have all an important role in updating legal frameworks, promoting big data practices and developing and incorporating tools into the big data design and practice that address societal concerns.

**Acknowledgements**

We thank the participants of the BYTE project focus groups and workshops for they insightful contributions.

This work has been funded by the European Commission through the BYTE project [FP7 GA 619551].

**References**

Bekkers, V., 2005. E-Government, Changing Jurisdictions and Boundary Management, in: Bekkers, V., Homburg, V. (Eds.), The Information Ecology of E-Government: E-Government as Institutional and Technological Innovation in Public Administration. IOS Press, pp. 53–71.

Blanke, T., 2014. Digital asset ecosystems: rethinking crowds and cloud. Chandos Publishing, Kidlington.

Boyd, D., Crawford, K., 2012. Critical questions for big data. Information, Commun. Soc. 15, 662–679. doi:10.1080/1369118X.2012.678878

Centre for Co-operation with European Economies in Transition, 1993. Glossary of industrial organisation economics and competition law. Organization for Economic Co-operation and Development.

Commonwealth of Australia, 2013. The Australian Public Services Big Data Strategy. Australian Government Information Management Office.

Cranor, L., Rabin, T., Shmatikov, V., Vadhan, S., Weitzner, D., 2016. Towards a Privacy Research Roadmap for the Computing Community.

Creswell, J.W., 2009. Research Design: Qualitative, Quantitative, and Mixed Methods Approaches, 3rd ed. Sage Publications.

Cuquet, M., Fensel, A., 2016. Big data impact on society: a research roadmap for Europe. arXiv 1610.06766.

Curry, E., 2016. The Big Data Value Chain: Definitions, Concepts, and Theoretical Approaches, in: Cavanillas, J.M., Curry, E., Wahlster, W. (Eds.), New Horizons for a Data-Driven Economy. Springer International Publishing, pp. 29–37. doi:10.1007/978-3-319-21569-3_3

Diebold, F.X., 2003. "Big Data" Dynamic factor models for macroeconomic measurement and forecasting. Adv. Econ. Econom. Theory Appl. Eighth World Congr. Econom. Soc. 32, 115–122.

Donovan, A., Finn, R., Oruc, S., Werker, C., Cunningham, S.W., Vega Gorgojo, G., Soylu,

A., Roman, D., Akerkar, R., Garcia, J.M., Lammerant, H., Galetta, A., De Hert, P., Grumbach, S., Faravelon, A., Ramirez, A., 2014. Report on legal, economic, social, ethical and political issues. doi:10.5281/ZENODO.49169

Dumbill, E., 2013. Making Sense of Big Data. Big Data 1, 1–2. doi:10.1089/big.2012.1503

Finn, R., Watson, H., Wadhwa, K., 2015. Exploring big "crisis" data in action: potential positive and negative externalities, in: Proceedings of the 12th International ISCRAM Conference.

Grumbach, S., Faravelon, A., Cuquet, M., Fensel, A., 2016. A roadmap for big data incorporating both the research roadmap and the policy roadmap.

Jagadish, H. V., Gehrke, J., Labrinidis, A., Papakonstantinou, Y., Patel, J.M., Ramakrishnan, R., Shahabi, C., 2014. Big data and its technical challenges. Commun. ACM 57, 86–94. doi:10.1145/2611567

Lammerant, H., 2016. Evaluating and Addressing the Impact of Big Data on Legal Frameworks, in: Building a European Digital Space, Proceedings of the 12th International Conference on Internet, Law & Politics. Barcelona, pp. 184–197.

Lammerant, H., De Hert, P., Lasierra Beamonte, N., Fensel, A., Donovan, A., Finn, R., Wadhwa, K., Grumbach, S., Faravelon, A., 2015a. Horizontal analysis of positive and negative societal externalities. doi:10.5281/ZENODO.104960

Lammerant, H., De Hert, P., Vega Gorgojo, G., Stensrud, E., 2015b. Evaluating and addressing positive and negative societal externalities. doi:10.5281/ZENODO.104962

Manovich, L., 2012. Trending: The promises and the challenges of big social data, in: Gold, M.K. (Ed.), Debates in the Digital Humanities. University of Minnesota Press, pp. 460–475.

Metcalf, J., Keller, E.F., Boyd, D., 2016. Perspectives on Big Data, Ethics, and Society.

Michael, K., Miller, K.W., 2013. Big Data: New Opportunities and New Challenges. Computer (Long. Beach. Calif). 46, 22–24. doi:10.1109/MC.2013.196

Pedreschi, D., Ruggieri, S., Turini, F., 2013. The Discovery of Discrimination, in: Custers, B., Calders, T., Schermer, B., Zarsky, T. (Eds.), Discrimination and Privacy in the Information Society. Springer Berlin Heidelberg, pp. 91–108. doi:10.1007/978-3-642-30487-3_5

Rossman, G.B., Rallis, S.F., 1998. Learning in the Field: An Introduction to Qualitative Research, 3rd ed. Sage Publications.

Rouvroy, A., 2016. "Of Data and Men" - Fundamental rights and freedoms in a world of massive data. Strasbourg.

Ruggieri, S., Pedreschi, D., Turini, F., 2010. DCUBE: discrimination discovery in databases, in: Proceedings of the 2010 International Conference on Management of Data - SIGMOD '10. ACM Press, New York, New York, USA, p. 1127. doi:10.1145/1807167.1807298

UN Global Pulse, 2012. Big Data for Development: Challenges & Opportunities.

Vega-Gorgojo, G., Donovan, A., Finn, R., Bigagli, L., Rusitschka, S., Mestl, T., Mazzetti, P., Fjellheim, R., Psarros, G., Drugan, O., Wadhwa, K., 2015. Case study reports on positive and negative externalities. doi:10.5281/ZENODO.166263

Vega-Gorgojo, G., Fjellheim, R., Roman, D., Akerkar, R., Waaler, A., 2016. Big data in the oil & gas upstream industry - a case study on the Norwegian continental shelf. Oil Gas Eur. Mag. 42, 67–77.

Yin, R.K., 2014. Case study research: design and methods. Sage Publications.

**Vitae**

**Dr. Martí Cuquet** is a postdoctoral researcher at the Semantic Technology Institute, Universität Innsbruck. He has a doctoral degree in Theoretical Physics from Universitat Autònoma de Barcelona (2012), a master degree in Photonics from Universitat Politècnica de Catalunya (2008) and a university degree in Physics from Universitat Autònoma de Barcelona (2007). His research has been at the crossroads of quantum information and network science, and more recently in big data and data science.

**Dr. Guillermo Vega-Gorgojo** is a postdoctoral researcher in the Department of Informatics at the University of Oslo in Norway. His research interests include big data, semantic web and visual search interfaces. Guillermo holds an associate professorship at the University of Valladolid, Spain.

**Hans Lammerant** is a PhD researcher at the Law, Science, Technology and Society research group (LSTS) at the Vrije Universiteit Brussel (VUB). He has studied philosophy (VUB) and law (Universiteit Antwerpen). His research focuses on the impact of new developments in data science and statistics on surveillance and privacy, and law in general.

**Dr. Rachel Finn** is a Practice Manager at Trilateral Research Ltd. Her research centres on the social impacts of new technology innovations, with a special focus on novel data practices, and she coordinates and manages multiple projects in this space. She has a PhD in Sociology from the University of Manchester (UK).

**Dr. Umair ul Hassan** is a postdoctoral Researcher at the Insight Centre of Data Analytics, National University of Ireland, Galway. His research interests include crowdsourcing, linked data, dataspaces, smart environments, and IT for social good. He is project manager for EU funded BYTE project at the National University of Ireland Galway. He is also a member of programme task force of Big Data Value Association.